\documentstyle[epsf,12pt]{article}

\catcode`\@=11
\def\numberbysection{\@addtoreset{equation}{section}
  \def\theequation{\arabic{section}.\arabic{equation}}}
\begin{document}
\addtolength{\baselineskip}{3.0mm}
\begin{titlepage}
\begin{flushright}{
    TIT/HEP-242/COSMO-39\\ May, 1994}
\end{flushright}
\begin{center}
  {\Large\bf Cosmological Sphaleron from Real Tunneling and Its Fate}\\
\vspace{4mm}
 {\bf Shuxue
   Ding}\def\thefootnote{\fnsymbol{footnote}}\footnote{Electronic address:
sding@th.phys.titech.ac.jp}

{ Department of Physics, Tokyo Institute of
    Technology,\\ Oh-Okayama, Meguro, Tokyo 152, Japan\\}

\vspace{8mm}
{\large ABSTRACT}
\end{center}
We show that the cosmological sphaleron of the Einstein-Yang-Mills
system can be produced from real tunneling geometries. The
sphaleron will tend to roll down to the vacuum or pure gauge
field configuration, when the universe evolves in the
Lorentzian signature region with the sphaleron and the
corresponding hypersurface being the initial data for the
Yang-Mills field and the universe, respectively. However, we can
also show that the sphaleron, although unstable, can be
regarded as a pseudostable solution because its lifetime is
even much greater than that of the universe.
\vspace{4mm}
\hfill\break
\noindent
PACS number(s): 98.80.Hw, 98.80.Cq.
\end{titlepage}
\clearpage \setcounter{section}{0}
\setcounter{equation}{0}
\numberbysection
\section{Introduction}

Since Bartnik and McKinnon \cite{BMcK} found a series of static
solutions of the Einstein-Yang-Mills equation, great interest
has been paid to the coupled system of the Yang-Mills field and gravity
\cite{VG,Bizon,SZ1,SZ2,GV,MW,BS,GS1,GS2,SW}. What is interesting
is that the Bartnik and McKinnon (BK) solutions were shown to be some
analogues of a sphaleron in the pure Yang-Mills theory or the
Weinberg-Salam model \cite{M,KM}. We call them EYM sphalerons.  This
means that the (Lorentzian) solutions sit on the top of the
potential barrier between topologically distinct vacua in the field
configuration space. If the collapse of BK solutions form black
holes with color, it will shed some light on the no-hair
theorem of black hole \cite{GS1}. Some authors also discussed
anomalous productions of fermions \cite{GS1,GS2}, with the
sphalerons as backgrounds.

The sphaleron solutions in the EYM system were also considered in
the context of cosmology \cite{GS2}. Previously several kinds of
cosmological solutions were found \cite{H,VD}; however, it has
recently been shown that some of the solutions are relevant to
sphalerons \cite{GS2}. One such sphaleron is a spacetime with
$S^3$ spatial section and homogeneous Yang-Mills field
\cite{GS2}.

In this paper we will discuss how and what kind of EYM sphalerons can
appear in the context of cosmology and how they develop later.
As we know a sphaleron is an unstable configuration which
corresponds to a saddle point on the energy functional, sitting
on the top of the energy barrier. If a spacetime has only a
Yang-Mills field as the source, classically, we cannot find a
mechanism to make the system on the top of the barrier statically
although the state exists theoretically, because the system always take the
least energy state (Of course here we are only concerned with the case
of zero-temperature, and we can say nothing about the existence
of such a mechanism at high temperature). However, as we
shall see, the cosmological sphaleron can be created by real
tunneling geometries, i.e., by the nucleation of the universe from
nothing represented by a Riemannian manifold with a single
totally geodesic boundary or by a quantum tunneling between two
Lorentzian signature regions which sandwich a Euclidean signature
region with the boundary surfaces being totally geodesic
\cite{HH,GH}. Such events are relevant to the WKB approximation
of the Wheeler-DeWitt equation.

In Sec. 2, we shall give a review of the decomposition of the
Einstein-Yang-Mills equation which is suitable for our purpose
. In Sec. 3, we will solve the constraint
equations with the real tunneling conditions both for the
spacetime and the Yang-Mills field configurations. In Sec. 4,
we discuss how the spacetime and the Yang-Mills field
configuration develop in the Lorentzian signature region with
the the real tunneling geometry and sphaleron as the initial
conditions. In Sec. 5 we draw some conclusions and
give some discussions.
\vspace{5mm}
\section{3+1 Decomposition of Einstein-Yang-Mills Equation}

Our aim is to discuss the time development of the universe and
Yang-Mills field configuration in the Lorentzian signature
region, with a real tunneling hypersurface $\Sigma$ as an
initial spatial section and with a purely magnetic Yang-Mills
field configuration on the hypersurface $\Sigma$ as initial
data. Here we only consider the Yang-Mills field with
$SU(2)$ as a gauge group, for simplicity. The results in the
papers can be generalized to any Yang-Mills field with arbitrary
compact gauge group.

As is well known, there are constraint equations in the
Einstein-Yang-Mills theory (the 3+1 decompositions of the EYM
equations are nontrivial \cite{SW,I,ADM}). The allowed initial data
are restricted to the constraint submanifold $\Sigma$ defined by
\cite{SW,I,ADM}
\begin{equation}
G^a \equiv 4[\sqrt{h}{\nabla}^i( E^a_i/{\sqrt{h}}) + {\epsilon}^{abc}
A^{bi}E^c_i] = 0,
\label{C1}
\end{equation}
\[
H_0 \equiv \sqrt{h}[-({}^3R - 2 \Lambda )+{1 \over h}({\pi}_{ik}{\pi}^{ik} - {1
\over 2}
{\pi}^2)] +
\]
\begin{equation}
 {\kappa \over 2}[({2 \over \sqrt{h}})E^a_i E^i_a +
\sqrt{h}F^a_{ik}F_a^{ik}] = 0,
\label{C2}
\end{equation}
\begin{equation}
H_i \equiv -2\sqrt{h} {\nabla}_k ({\pi}_i^k/{\sqrt{h}}) + 2
\kappa F^a_{ik}E^k_a = 0.
\label{C3}
\end{equation}
at each point $x\in \Sigma$ \cite{I,SW,ADM}.
Here $E^a_i= \sqrt{h} F^a_{i0} $ is the electric field of the
Yang-Mills field, ${\nabla}_k$ is the covariant derivative
operator on $\Sigma$ compatible with the 3-metric $h_{ik} (i,k =
1,2,3)$, and ${}^3R$ denotes the scalar curvature with respect
to $h_{ik}$. $\kappa$ stands for $8 \pi G$, the coupling
constant between matter field and gravity.

As a matter of fact, the constraint (\ref{C1}) is just the
orthogonal component of the Yang-Mills field equation
\begin{equation}
(*D*F)_{\bot} = 0,
\end{equation}
in the formalism of differential form, where the asterisk
denotes is the Hodge dual.

Here the metric is decomposed as
\begin{equation}
ds^2=-(N^2-N^iN_i)dt^2 + 2 N_idt dx^i + h_{ik}dx^i dx^k,
\label{M}
\end{equation}
in the Lorentzian signature region. ${\pi}^{ik}$ is the momentum
canonically conjugate to $h^{ik}$ which is related to the
extrinsic curvature $K^{ik}$ of $\Sigma$,
\begin{equation}
{\pi}^{ik} = \sqrt{h}(K^{ik} - h^{ik} K).
\label{Mom}
\end{equation}
 The extrinsic curvature is defined as
\begin{equation}
K_{ik} = {1 \over 2N}(h_{ik,0} - {\nabla}_kN_i - {\nabla}_iN_k),
\label{Ex}
\end{equation}
where $K$ is the trace of $K_{ik}$.

The evolution equations of geometry and Yang-Mills field are
given by
\begin{equation}
{\cal L}_0 E^a_k = N[D^i F^a_{ik} + {1 \over 2} K E^a_k + E^b_k
{\epsilon}^{abc} A^c_0] + {\nabla}^i N F^a_{ik} +
{\cal L}_{\bf N} E^a_k,
\label{E1}
\end{equation}
\[
{\cal L}_0 K^i_k =N[ R^i_k + K K^i_k + 2 {\kappa}(E^{ai} E^a_k +
F^{ai}_l F^{al}_k) - \kappa {\delta}^i_k(E^{al} E^{al}
\]
\begin{equation}
- {1 \over 2} F^a_{lm} F^{alm})] -
{\nabla}^i{\nabla}_k N + {\cal L}_{\bf N} K^i_k.
\label{E2}
\end{equation}
Here $D^i$ stands for the covariant derivative compatible both with
Yang-Mills and spacetime connections and $\cal L$ stands for the
Lie derivative. The constraint equation (\ref{C1}) together
with the evolution equation (\ref{E1}) are equivalent to the Yang-Mills
equation. Similarly the constraint equations (\ref{C2}), (\ref{C3})
and the evolution equation (\ref{E2}) are equivalent to the Einstein
equation. It is worthy of noticing that the evolution equation
(\ref{E1}) is the $\Sigma$ component of the Yang-Mills field equation
\begin{equation}
(*D*F)_{\Sigma} = 0.
\end{equation}
in the formalism of differential form.
\vspace{5mm}
\section{ Real Tunneling Geometry and Yang-Mills Field}
Real tunneling geometries arise in the WKB approximation for
the solution of the Wheeler-DeWitt equation as a special case
of the general situation, in which one considers ``complex
paths'' which are spacetimes with complex metrics. Real
tunneling geometries are partially Lorentzian and partially
Riemannian \cite{HH,GH}, as a special complex geometry.
Alternatively the Lorentzian and Riemannian portions may be
regarded as different real slices of a complex spacetime $M_c$,
which have a common boundary $\Sigma$ (spacelike). This
hypersurface $\Sigma$ acts as an initial Cauchy surface for the
Lorentzian spacetime and the matter field on it as the initial
matter field of the Lorentzian spacetime.

It was shown \cite{HH,GH} that the extrinsic curvature $K_{ik}$
and matter field must be continuous across $\Sigma$ to make the
action finite. The continuity of these require that on $\Sigma$
\begin{equation}
K_{ik} = 0,
\label{TG}
\end{equation}
and momentum of the matter field vanishes, to our case
\begin{equation}
E^a_k = 0.
\label{TC}
\end{equation}

Let us find a solution which satisfies the above conditions and
constraint equations (\ref{C1}), (\ref{C2}), and (\ref{C3}) on
the initial hypersurface $\Sigma$. Substituting the conditions
(\ref{TG}) and (\ref{TC}) into the constraints (\ref{C1}),
(\ref{C2}), and (\ref{C3}), we find the first and the third
constraints are automatically satisfied, and the Hamiltonian
(\ref{C2}) constraint of gravity becomes
\begin{equation}
^3 R = {\kappa \over 2}F^a_{ik} F^{ik}_a + 2 \Lambda
\label{C2P}
\end{equation}

This can be regarded as a constraint equation for the curvature
of spatial section if we are given a pure magnetic field. The
simplest but nontrivial situation will be the spatial section
with a constant curvature. Because, however, the right hand side
of the equation (\ref{C2P}) is always positive if the cosmological
constant is positive, there exists only a positive constant
curvature solution, i.e., $S^3$, for spatial section. For
simplicity of discussions, we introduce a four-dimensional
Euclidean space with the induced metric
\begin{equation}
d{\Omega}^3 = (dx^1)^2 + (dx^2)^2 + (dx^3)^2 +(dx^4)^2,
\label{MS}
\end{equation}
for a three-dimensional sphere $S^3$,
\begin{equation}
(x^1)^2+(x^2)^2 +(x^3)^2 +(x^4)^2=1.
\label{S3}
\end{equation}

Because there exists a natural imbedding of $S^3$ into the group
manifold $SU(2)$, $g = x^4 + i x^i {\sigma}^i$, the left
invariant one-form of the gauge group is given by
\begin{equation}
\mbox{\boldmath$e$}^i=2(x^4 dx^i - x^i dx^4 + {\epsilon}_{ijk} x^j dx^k), \ \ \
i=1,2,3.
\label{L}
\end{equation}
Here the one-form $\mbox{\boldmath$e$}^i$ obeys the Maurer-Cartan
structure equation
\begin{equation}
d\mbox{\boldmath$e$}^i = {1 \over 2}
{\epsilon}_{ijk}\mbox{\boldmath$e$}^j \wedge
\mbox{\boldmath$e$}^k.
\label{MC}
\end{equation}

There is a natural ansatz for the gauge field on $S^3$ which takes the form
\begin{equation}
A=-{1 \over e} f \mbox{\boldmath$e$}^i {{\sigma}^i \over 2},
\label{PO}
\end{equation}
where $f$ is a constant to be determined. The field strength associated with
(\ref{PO}) is
\begin{equation}
F= {1 \over 2e}f(f-1){{\epsilon}^i}_{jk} {\mbox{\boldmath$e$}}^j \wedge
{\mbox{\boldmath$e$}}^k {{\sigma}^i \over 2}.
\label{Field}
\end{equation}
This is a kind of homogeneous Yang-Mills field.
{}From the expression (\ref{Field}) or directly from (\ref{PO}),
one observes that $f=0$ and $f=1$ correspond to vacuum or
pure gauge field configurations.

To get a complete solution with a given initial value on the
hypersurface $\Sigma$, only considering the constraint equations
is not enough, and we have to take advantage of the evolution
equations. In terms of differential form, the evolution equation
(\ref{E1}) becomes
\begin{equation}
({\partial}_0 E^a_k){\mbox{\boldmath$e$}}^k {{\sigma}^a \over 2} =
[ {\tilde{\mbox{\boldmath$e$}}}^l F^a_{lk} - {1 \over 2}
{{\epsilon}_k}^{ij} F^a_{ij}
 + e {\epsilon}^{abc}( A^{aj} F^b_{jk} - A^a_0
E^b_k)]{\mbox{\boldmath$e$}}^k {{\sigma}^a \over 2},
\label{E1P}
\end{equation}
where
\begin{equation}
 {\tilde{\mbox{\boldmath$e$}}}^i = {1 \over 2} (x^4 {\partial \over
  \partial x^i} - x^i {\partial \over \partial x^4} +
 {{\epsilon}^i}_{jk} x^j {\partial \over \partial x^k}), \ \ \
\end{equation}
is the left invariant vector field dual to (\ref{L}).
By the real tunneling condition (\ref{TC}) together with our
ansatz (\ref{PO}) and the field strength (\ref{Field}), the
evolution equation for the Yang-Mills field now becomes
\begin{equation}
{1 \over e} f (2f-1) (f - 1) {\epsilon}_{ijk}
{ \mbox{\boldmath$e$}}^j \wedge {\mbox{\boldmath$e$}}^k
{{\sigma}^i \over 2} = 0.
\label{Equ}
\end{equation}

We immediately notice that, $f=1/2$ is a nontrivial static
solution of the equation (\ref{Equ}). The Chern-Simons index is
easy to be calculated, which is
\begin{equation}
N_{CS}= 3 f^2  [1 - (2/3) f ] = 1/2.
\label{CS}
\end{equation}
Then this static solution is a sphaleron.

At this initial moment, the universe is just equivalent to the
Hosotani type universe \cite{H}. However we shall see in the
next section that the spacetime and Yang-Mills field will evolve
to make the universe leave the Hosotani universe due to the
instability of the sphaleron.

The Yang-Mills field configuration, i.e., the sphaleron, is
corresponding to a local
maximum point in the viewpoint of the Lorentzian signature region,
while it is corresponding to a local minimum point in the
viewpoint of Euclidean signature region \cite{C}. It is just
from this reason that, we can extend the solution to the whole
Euclidean signature region, i.e., $ F^i = - {1 \over 8e} {{\epsilon}^i}_{jk}
{\mbox{\boldmath$e$}}^j \wedge {\mbox{\boldmath$e$}}^k
{\sigma}^i/2 $. Our spacetime is endowed with the metric
\begin{equation}
ds^2 = a_0^2 ( d{\tau}^2 + d{\Omega}^3) = a_0^2 (- d{\eta}^2 + d{\Omega}^3),
\label{MF}
\end{equation}
where $a_0^2$ is any constant having dimension $[L^2]$, $\tau$
is the Euclidean time and $\eta$ is the conformal time in the
Lorentzian signature region, which have a relation with the
comoving time $t$ defined in the next section. A Wick rotation $\tau =
i\eta$ will act on the initial hypersurface $\Sigma$ because the
hypersurface may be defined as the time slice with $\tau = \eta
= 0$. With the viewpoint of the Euclidean signature region, the
hypersurface $\Sigma$ corresponds to the Yang-Mills wormhole
\cite{HO} at the throat.
\vspace{5mm}
\section{ Evolution in Lorentzian Signature Region}
Let us discuss how the spacetime and the Yang-Mills field develop
with (\ref{PO}), (\ref{Field}), and (\ref{MF}) being the initial
data. It is reasonable to suppose that the spatial section still has
the symmetry of $S^3$ and the Yang-Mills field homogeneously
distributes in the spatial section $\Sigma$ at a later time,
because there is no mechanism to break such a symmetry in our
model. However, the radius of $\Sigma$ and then $f$ in the
ansatz (\ref{PO}) will change in time. Then following Gibbons
and Steif \cite{GS2}, the ansatz for the Yang-Mills connection
one-form is
\begin{equation}
A=-{1 \over e} f(\eta)\mbox{\boldmath$e$}^i {{\sigma}^i \over
  2}.
\label{POT}
\end{equation}
 This will make the electric field nonzero,
\begin{equation}
F= F^i{{\sigma}^i \over 2}, \ \ \ F^i= -{{\dot f} \over e}
d{\eta}\wedge{\mbox{\boldmath$e$}}^i + {1 \over
  2e}f(f-1){{\epsilon}^i}_{jk} {\mbox{\boldmath$e$}}^j \wedge
{\mbox{\boldmath$e$}}^k.
\label{FieldT}
\end{equation}
The metric is supposed to be
\begin{equation}
ds^2 = a^2(\eta) ( -d{\eta}^2 + d{\Omega}^3) = -dt^2 + a^2(t)d{\Omega}^3,
\label{MFT}
\end{equation}
based on the symmetry of the spacetime. Here the cosmic time $t$ and
the conformal time $\eta$ are related by
${\partial}\eta/\partial t = a^{-1}$.

We will solve the constraint equations and evolution equation
with the initial values
\begin{equation}
f( 0 ) = {1 \over 2}, \ \ \ f^{\prime}( 0 ) = 0,
\label{I1}
\end{equation}
and
\begin{equation}
a^2( 0 ) = a^2_0.
\label{I2}
\end{equation}

We first consider the constraints (\ref{C1}), (\ref{C2}), and
(\ref{C3}). Again the constraints (\ref{C1}) and (\ref{C3}) are
automatically satisfied by the ansatz (\ref{POT}) and
(\ref{MFT}).

{}From the constraint (\ref{C2}), with (\ref{POT}),
(\ref{FieldT}) and (\ref{MFT}), we obtain
\begin{equation}
{(\dot{a})^2 \over a^2}= {\Lambda \over 3} a^2 + 4 \kappa e^{-2}
[{1 \over 2}(\dot{f})^2 + 2f^2 ( f -1)^2]{1 \over a^2} - 1.
\label{C2S}
\end{equation}

Now let us turn to the evolution equations (\ref{E1}) and
(\ref{E2}). By the ansatz (\ref{MFT}), we have $N = a(\eta)$ and
$ N_i = 0$. The evolution equation (\ref{E1}) is relevant to
the developing of the Yang-Mills field. Because the Yang-Mills
equation is conformal invariant, we can discuss the equation
(\ref{E1}) in a conformally transformed metric $ ds^{\prime 2} =
-d{\eta}^2 + d{\Omega}^2$, i.e., $N^{\prime} = 1$ and $
N^{\prime}_i =0 $. The evolution equation (\ref{E1}) can be
simplified into the form
\begin{equation}
{ d^2 f \over d{\eta}^2} = - {\partial V \over
  \partial f}, \ \ \ V(f) = 2 f^2 (f - 1)^2,
\label{C1E}
\end{equation}
or
\begin{equation}
{1 \over 2} {\dot{f}}^2 + V(f) = E,
\label{C1EP}
\end{equation}
where $E$ is a constant to be determined.

The evolution equation (\ref{E2}) can be written as
\begin{equation}
a \ddot{a} + 2 {\dot{a}}^2 = \kappa e^2 a^{-2} + \Lambda a^2 -
2.
\label{C2E}
\end{equation}

The general solution of the equation (\ref{C1EP}) can be written
as an elliptic function. However, from the initial condition
(\ref{I1}), we have $E=1/8$. For such a special case the
solution of the equation is simply given by
\begin{equation}
f(\eta) = {1 \over 2} \pm {\sqrt{2} exp(\sqrt{2} {\eta})
  \over 1 + exp(2\sqrt{2} {\eta})}.
\label{YS}
\end{equation}
This solution was given by Gibbons and Steif \cite{GS2}, taken as
a special example for a rolling down cosmological sphaleron.
However, we give it as an inevitable result if the sphaleron comes
from real tunneling.

We can see that at $\eta = - \infty$, the Yang-Mills field becomes the
sphaleron configuration, while as $\eta \rightarrow {\sqrt{2}
  \over 2} ln( \sqrt{2} - 1 )$ the Yang-Mills field approaches the
vacuum or pure gauge configuration. From the equation
(\ref{YS}), we see that the lifetime of the sphaleron to roll
down is infinite.

By the equation (\ref{C1EP}) with $E=1/8$, the equation (\ref{C2S}) can be
simplified as
\begin{equation}
a^{-2} (\dot{a})^2 = {\Lambda \over 3} a^2 + {1 \over 2} \kappa e^{-2} a^{-2} -
1.
\label{C2SC}
\end{equation}
which is independent of the specific form of $f(\eta)$. To
contrast this with a Newtonian particle moving on a potential,
we can write the equation (\ref{C2SC}) as
\begin{equation}
{1 \over 2}(\dot{a})^2 + U(a) = E_a
\label{Cl}
\end{equation}
where $E_a = {\kappa \over 4e^2}$ is a constant analogous to
the energy  and $U(a) ={1 \over 2} (a^2 - {\Lambda \over 3}
a^4)$ the potential of classical particle.
When the cosmological constant $\Lambda$ vanishes, the
``potential'' is simply ${ 1 \over 2} a^2$ and the solution can
easily be got
\begin{equation}
a^2(\eta) = {\kappa \over 2e^2} sin^2( \eta + {\eta}_0)= {\kappa
  \over 2e^2} ( 1
- ( 1 - {t \over \sqrt{{\kappa \over 2e^2}}})^2).
\label{AS0}
\end{equation}

{}From the equation (\ref{AS0}), we see the lifetime of the universe
is $\Delta \eta = \pi$. However, from the equation
(\ref{YS}), the lifetime of the sphaleron to roll down to the
vacuum or pure gauge configuration is, infinite, much longer than the
lifetime of the universe. As we know a sphaleron solution is an
unstable solution generally. However, for our case the sphaleron
solution can be regarded as a pseudostable solution because it
will roll down very slowly compared with the evolution of the
universe.

As is well known, a cosmological constant, if it exists and is
positive, is always likely to drive the universe to expand longer.
Now let us see if a sphaleron can complete a whole process of
rolling down to the vacuum or pure gauge configuration, provided
that $\Lambda > 0$. For this sake, we solve the equation (\ref{C2SC})
and change it into the form
\begin{equation}
{da \over d\eta} = \sqrt{\Lambda \over 3} \sqrt{(a^2 - A)(a^2 -
  B)}.
\label{EL}
\end{equation}
Here
\[
A = {3 \over 2 \Lambda}(1 + \sqrt{1 - {2 \kappa \Lambda
    \over 3 e^2}})
\]
and
\[
B = {3 \over 2 \Lambda}(1 - \sqrt{1 -
  {2 \kappa \Lambda \over 3 e^2}})
\]
are two constants.

The solution of the equation (\ref{EL}) can be written as an elliptic
function generally. If the universe initiates from the
hypersurface with radius $a_0 < \sqrt{A}$, the situation is
similar to those when the cosmological constant is vanishing. If
the universe initiates from a
hypersurface with $a_0 > \sqrt{A}$, the universe will expand
forever. By the real tunneling, it is possible for a universe to
start from a finite region \cite{V}, and what is more the
potential $U(a)$ in the equation (\ref{C1}) is very similar to
those discussed by Blau $et al$. with a domain wall emerging from a
finite size by quantum tunneling \cite{BGG}. In this case the
solution for the equation (\ref{EL}) is
\begin{equation}
a(\eta) = \sqrt{A} dc(\sqrt{A \Lambda \over 3} (\eta - {\eta}_0)
| {B \over A}),
\label{SS}
\end{equation}
where $dc(x | k)$ is one of the elliptic functions \cite{M-T}.
Here ${\eta}_0$ is a constant determined by
\begin{equation}
a_0 = a(\eta=0) = \sqrt{A} dc(\sqrt{A \Lambda \over 3} (- {\eta}_0)
| {B \over A}).
\label{SSC}
\end{equation}

As a matter of fact, the solution of the equation (\ref{C2SC})
can be gotten more easily in the coordinate $t$,
\begin{equation}
a^2(t) = {3 \over \Lambda c}[{\kappa \Lambda \over 3 e^2} -
{1 \over 4} (1 - c\cdot exp(- 2 \sqrt{{\Lambda \over 3}t}))^2 ]
exp(2\sqrt{{\Lambda \over 3}t}).
\label{SST}
\end{equation}
Here $c$ is a constant determined by the initial condition
(\ref{I2}),
\begin{equation}
c = [1 - {2\Lambda \over 3} a^2_0 + \sqrt{ (1 - {2\Lambda
    \over 3} a_0^2 )^2 + {2 \kappa \Lambda \over 3 e^2} -1 } ].
\end{equation}

Because the solution for the Yang-Mills field configuration
(\ref{YS}) is written as a function of $\eta$, we have to
compare the lifetimes for the sphaleron to roll down and the
universe to develop by $\eta$.
By the equation (\ref{SS}), the scale factor $a$ will get
infinity when $\eta \rightarrow 2B/A$. That is to say,
although the universe may expand to infinity, the lifetime is
finite in the coordinate of $\eta$. Because the rolling from the
sphaleron will take infinite time ($\eta$), this whole
procedure cannot be finished within one period of the
evolution of the universe. In this case the sphaleron can also be
regarded as a pseudostable solution.
\vspace{5mm}
\section{Conclusions and Discussions }

We have obtained the cosmological Yang-Mills field which
describes the rolling down from the cosmological sphaleron in
the potential picture. The cosmological sphaleron is produced by
a real tunneling geometry from nothing or from another
Lorentzian signature region emerging from an Euclidean signature
region. It is natural to ask such a question as how probable
the real tunneling is? To answer this
question, the amplitude for a hypersurface to emerge from a real
tunneling with sphaleron on it should be calculated,
\begin{equation}
\Psi(S^3, Sphaleron)= N e^{-S_E} = N,
\label{Am}
\end{equation}
because the Euclidean action vanishes for the sphaleron states.
Here N is a prefactor in WKB approximation which in principle calculable.
This amplitude is a local maximum point because
the action is a local minimum point in the Euclidean signature region.
We can see this fact from the equation (\ref{C1E}) by making the
Wick rotation, $\eta = -i\tau$. The potential becomes $-2f^2(f - 1)$,
and $f = 1/2$ is a minimum value of it. As a matter of fact,
this is very similar to the discussion for the BK solutions
corresponding to saddle points for the Euclidean energy functional
\cite{MW}.

After the real tunneling, the solution for $f(t)$ [or
$f(\eta)$] and $a^2(t)$ still describes a homogeneous Yang-Mills
field and the homogeneous and isotropic universe with $S^3$ as a
spatial section $\Sigma$. The electrical field is produced
during the rolling and evolution for the Yang-Mills field and
the universe, respectively.

\vspace{5mm}
\vspace{8mm}
{\large{\bf Acknowledgments}}
\par
The author would like to express his sincere thanks to Professor
A. Hosoya for some helpful discussions and careful reading of the
manuscript. He is also grateful to Dr. Y. Fujiwara and Dr. M.
Siino for useful discussions. This work was supported in part by
the Japanese Government.
\eject


\eject


\end{document}